\documentclass[a4paper, 11pt]{article}

\usepackage[margin=20mm, top=20mm]{geometry}
\usepackage{graphicx}
\usepackage{xcolor}

\usepackage{amsmath,amssymb,bm}
\usepackage{amsthm}
\usepackage{braket}
\usepackage{afterpage}
\usepackage{authblk}
\usepackage{here}
\usepackage{cite}

\usepackage{caption}
\usepackage[subrefformat=parens]{subcaption}
\numberwithin{equation}{section}

\let\oldchi\chi
\renewcommand{\chi}{\raisebox{0.4ex}{$\oldchi$}}

\usepackage{fancyhdr}

\usepackage{hyperref}
\hypersetup{%
setpagesize=false,%
bookmarksdepth=2,%
bookmarksnumbered=true,
colorlinks=true,%
linkcolor=blue,%
citecolor=blue,%
urlcolor=blue,%
pdftitle={},%
pdfsubject={},%
pdfauthor={},%
}

\allowdisplaybreaks[1]

\title{Tensor renormalization group approach to entanglement entropy}
\date{\today}
\author[1]{Takahiro H{\sc ayazaki}\footnote{email: t\_hayazaki@hep.s.kanazawa-u.ac.jp}}
\author[2]{Daisuke K{\sc adoh}\footnote{email: kadoh@mi.meijigakuin.ac.jp}}
\author[1]{Shinji T{\sc akeda}\footnote{email: takeda@hep.s.kanazawa-u.ac.jp}}
\author[2]{Gota T{\sc anaka}\footnote{email: gotanak@mi.meijigakuin.ac.jp}}
\affil[1]{\it Institute for Theoretical Physics, Kanazawa University, Kanazawa 920-1192, Japan}
\affil[2]{\it Institute for  Mathematical Informatics, Meiji Gakuin University, Kanagawa 244-8539, Japan}

\begin{document}
\maketitle
\thispagestyle{fancy} 
\fancyhead{}           
\fancyhead[R]{KANAZAWA-25-03}
\renewcommand{\headrulewidth}{0pt}

\begin{abstract}
	We propose a method to compute the entanglement entropy (EE) using the tensor renormalization group (TRG) method.
The reduced density matrix of a $d$-dimensional quantum system is represented as a $(d+1)$-dimensional tensor network.
We develop an explicit algorithm for $d=1$ that enables the calculation of EE for single-interval subsystems of arbitrary size.
We test our method in two-dimensional tensor network of the Ising model.
The central charge is obtained as $c=0.49997(8)$ for $D=96$, which agrees with the theoretical prediction within an error, demonstrating the accuracy and reliability of our proposed method.
\end{abstract}
\section{Introduction}
Quantum entanglement is a phenomenon which does not appear in classical mechanics, and is important to understand quantum characteristics of field theory.
Previous studies have investigated relations between quantum entanglement and various topics, such as quantum phase transition\cite{Osborne:2002zz, Vidal:2002rm}, quantum information \cite{Nielsen_Chuang_2010}, and quantum gravity\cite{Hawking:1975vcx, VanRaamsdonk:2010pw, PhysRevD.14.2460}.
Entanglement entropy (EE) is a kind of von Neumann entropy that measures quantum entanglement between two subsystems into which a quantum system is divided.
Many studies that quantitatively evaluate EE to investigate quantum entanglement have been performed.

It is a challenging task to compute EE, especially for strongly coupled quantum field theories.
A few analytical results are obtained in limited theories, such as low-dimensional conformal field theories \cite{Cardy:2007mb, Calabrese:2009qy} and some cases related with holography \cite{Ryu:2006ef}.
EE has also been studied numerically using the Monte Carlo method \cite{Buividovich:2008kq,Nakagawa:2009jk,Nakagawa:2010kjk,Itou:2015cyu,Rabenstein:2018bri,Bulgarelli:2023ofi,Jokela:2023rba}.
In this case, EE is obtained through the $n$-th R\'enyi entropy with the replica trick and an additional extrapolation $n\to 1$.
It would be beneficial to develop a method to compute EE without the extrapolation and for general theories with sign problems, where the Monte Carlo method faces difficulties.

The tensor renormalization group (TRG) method \cite{Levin:2006jai} is another numerical approach that can be applied to evaluate EE.
This method is free from sign problems, and the density matrix that defines EE can be evaluated without the replica trick.
{
The original TRG algorithm was developed for two-dimensional spin systems, and later applied to quantum field theories \cite{Shimizu:2012wfa,Kawauchi:2016dcg, Hirasawa:2021qvh} and extended to higher-dimensional theories using the Higher-order TRG (HOTRG) algorithm \cite{Xie:2012mjn} and several improved algorithms \cite{Adachi:2019paf, Kadoh:2019kqk, Nakayama:2023ytr}.
The computation of EE with the TRG method has been mainly investigated for the half-space \cite{PhysRevB.89.075116,Yang:2015rra,Bazavov:2017hzi,Luo:2023ont}.}

In this paper, we propose a method to evaluate the EE for arbitrary subsystems given by a single interval which is not limited to half space.
We test our method in 2d classical Ising model, demonstrating its effectiveness and accuracy.
This paper is organized as follows.
In section \ref{sec:theory}, we explain how to compute the entanglement entropy of $d$-dimensional quantum systems using the TRG method for $(d+1)$-dimensional tensor networks.
We then present our method to compute the entanglement entropy of arbitrary single-interval subsystems in one-dimensional quantum systems.
The numerical tests are performed in 2d classical Ising model in section \ref{sec:result}.
The last section is devoted to give a summary and outlook.
In appendix A, the notational details are given.

\section{Theory}
\label{sec:theory}

\subsection{Tensor network representation of the quantum many-body systems}
We consider a quantum system on a $d$-dimensional lattice with a local Hamiltonian $\hat{H}$.
For given density matrix $\rho$, the entanglement entropy of a subsystem $A$ is defined as
\begin{align}
	S_A = - \mathrm{Tr} ( \rho_A \log (\rho_A) ),
\label{eqn:EE_SA}
\end{align}
where $\rho_A = \mathrm{Tr}_{\bar{A}}(\rho)$ is the reduced density matrix for $A$, and $\bar{A}$ is the complement of $A$.
The density matrix $\rho$ may be represented as a $(d+1)$-dimensional tensor network.
We briefly explain this point here.
See Appendix~\ref{sec:notation} for the detailed notations of the tensor network.

\begin{figure}[ht]
	\centering
	\begin{minipage}[b]{0.48\linewidth}
		\centering
		\includegraphics[keepaspectratio, scale=0.5]{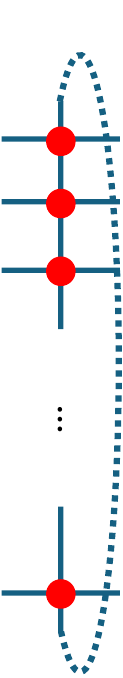}
		\subcaption{$e^{-\Delta \beta \hat{H}}$ for $d=1$.}
		\label{fig.time_evolution_tensor_network}
	\end{minipage}
	\begin{minipage}[b]{0.48\linewidth}
		\centering
		\includegraphics[keepaspectratio, scale=0.5]{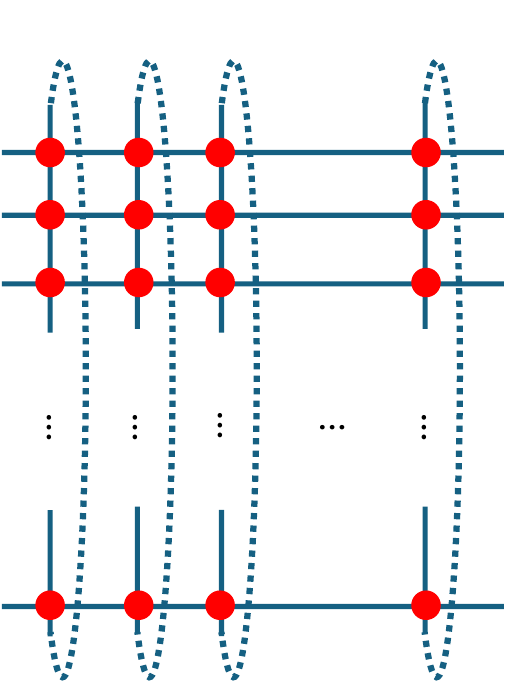}
		\subcaption{$e^{- \beta \hat{H}}$ for $d=1$.}
		\label{fig.time_evolution_tensor_network_finite_time}
	\end{minipage}
	\caption{ Tensor network representation of $e^{-\Delta \beta \hat{H}}$ and $e^{-\beta \hat{H}}$ for $d=1$.
		Contraction of indices in the vertical direction (the dotted lines) corresponds to the periodic boundary condition in spatial direction.}
	\label{fig.time_evolution_tensor_networks}
\end{figure}

The density matrix $\rho$ of the Gibbs state is defined as
\begin{align}
	\rho = \frac{e^{-\beta \hat{H}}}{\mathrm{Tr}(e^{-\beta \hat{H}})},
	\label{eq.density_matrix}
\end{align}
where $\beta$ is the inverse temperature.
The Boltzmann factor $e^{- \beta \hat{H}}$ includes nonlocal terms like $\hat{H}^2, \hat{H}^3, \dots$,
even if $\hat{H}$ is local.
However, for an infinitesimally small $\Delta\beta$,
higher order terms can be neglected as $e^{- \Delta \beta \hat{H}} \simeq 1 - \Delta\beta \hat{H}$.
Since $\hat{H}$ is a local operator acting on each spin variable, $e^{- \Delta \beta \hat{H}}$ is represented as a locally connected transformations and may be represented as a tensor network as Fig.~\ref{fig.time_evolution_tensor_network} for $d=1$.
Its tensor components are determined by parameters of $\hat{H}$.
As $e^{- \beta \hat{H}}=\lim_{N\to\infty}(e^{-\frac{\beta}{N}\hat{H}})^N$,
the Boltzmann factor $e^{- \beta \hat{H}}$ for finite $\beta$ is $N$ copy of Fig.~\ref{fig.time_evolution_tensor_network}, which is shown in Fig.~\ref{fig.time_evolution_tensor_network_finite_time}.
The density matrix \eqref{eq.density_matrix} can thus be represented as a tensor network (Fig.~\ref{fig.time_evolution_tensor_network_finite_time}).
Each external line in Fig.~\ref{fig.time_evolution_tensor_network} and \ref{fig.time_evolution_tensor_network_finite_time} corresponds to the local degrees of freedom of the quantum system.
Partial trace over a subsysmtem is performed by contracting the corresponding indices.

\begin{figure}[ht]
	\centering
		\includegraphics[keepaspectratio, scale=0.5]{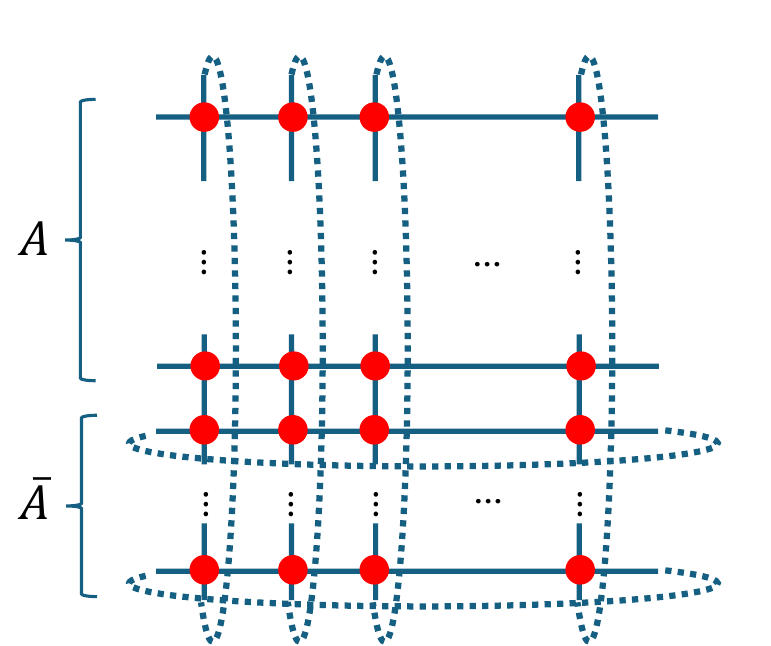}
		\caption{Tensor network representation of $\rho_A$ in $d=1$.}
		\label{fig.reduced_density_matrix_tensor_network}
\end{figure}

Consider $d=1$ and take a single interval as the subsystem $A$.
For that case, Fig.~\ref{fig.reduced_density_matrix_tensor_network} is a tensor network representation of $\rho_A$.
The indices associated with $\bar{A}$ are contracted and exhibit periodicity in the imaginary time direction, while the indices associated with $A$ remain open and form the matrix indices of $\rho_A$.
For arbitrary $d$ and $A$, the reduced density matrix is represented as a $(d+1)$-dimensional tensor network with closed external lines corresponding to $\bar{A}$.

It is important to note that these tensor networks (Fig.~\ref{fig.time_evolution_tensor_networks} and \ref{fig.reduced_density_matrix_tensor_network})
are given as a locally connected homogeneous network of a single tensor represented by a red dot.
This is because the Hamiltonian $\hat{H}$ is local and invariant under the spatial translation.
The path integral of the lattice field theory can be represented as a tensor network of similar structure.

The density matrix of a ground state $\ket{\psi}$,
\begin{align}
	\rho = \ket{\psi}\bra{\psi}
	\label{eq:ground_state_density_matrix}
\end{align}
characterizes the system at $T=0$ where quantum phase transition may occur.
The reduced density matrix and the entanglement entropy of the ground state are obtained from the zero temperature limit of \eqref{eq.density_matrix}.
The critical behavior of $d$-dimensional quantum system at $T=0$ is equivalent to that of a $(d+1)$-dimensional classical model at finite temperature \cite{suzukiRelationshipDDimensionalQuantal1976}.
In spin systems, although the the quantum Hamiltonian corresponds to an anisotropic model, the isotropic model and its tensor network are also used to study EE \cite{PhysRevB.89.075116,Yang:2015rra}.

The premise of our method is that the reduced density matrix $\rho_A$ is expressed as a tensor network such as Fig.~\ref{fig.reduced_density_matrix_tensor_network}.
In that case, the entanglement entropy can be computed for any subsystem size.

\subsection{\texorpdfstring{Our method to compute $S_A$ of arbitrary subsystem $A$}{Our method to compute S\_A of arbitrary subsystem A}}
We focus on the one-dimensional quantum system where the total system is divided into two intervals $A$ and $\bar{A}$, and assume that $\rho_A$ is given as a tensor network
as shown in Fig.~\ref{fig.reduced_density_matrix_tensor_network}.
The tensor is denoted by $T_{ijk\ell}$, where all indices $i,j,k,\ell$ run from $1$ to $D$.
Here, $L, N$, and $\ell$ represent the sizes of the spatial direction, the imaginary time direction, and the subsystem $A$, respectively.
The presented method is applicable to any subsystem size $\ell$.
For notational details, see Appendix~\ref{sec:notation} .

\begin{figure}[H]
	\centering
	\includegraphics[keepaspectratio, scale=0.7]{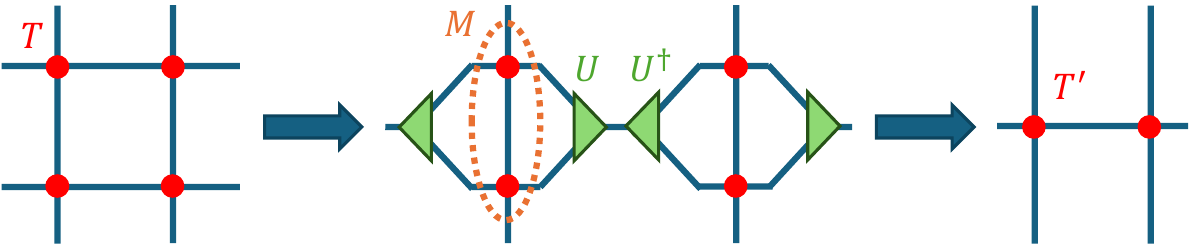}
	\caption{HOTRG algorithm for the two-dimensional tensor network.
		The tensor $M$ is obtained by contracting two tensor $T$s inside dotted loop.
		The isometry matrix $U$ is a set of eigenvectors of the matrix $M^\dagger M$ corresponding to $D$ largest eigenvalues.}
	\label{fig.HOTRG}
\end{figure}

Our method is based on the HOTRG algorithm \cite{Xie:2012mjn} in which two adjacent tensors are coarse-grained into a single tensor $T'$
as shown in Fig.~\ref{fig.HOTRG}.
The left panel shows a part of the whole network,
and the center panel shows a renormalization process.
Let $M$ be the $D^4 \times D^2$ matrix obtained by contracting two tensors $T$ inside the dotted loop.
The isometry matrix $U$ is constructed from the eigenvectors of $M^\dagger M$ corresponding to the $D$ largest eigenvalues.
Since $U$ is a part of the unitary matrix that diagonalizes $M^\dagger M$,
we have $UU^\dagger \neq I$ and $U^\dagger U = I$.
In Fig.~\ref{fig.HOTRG}, $UU^\dagger$ is inserted into the network,
and the value of the network is approximated.
The network in the left panel is renormalized into a network of $T'$ defined as $T'= U^\dagger M U$.
Since $T'$ is made of two $T$s, each renormalization reduces the number of tensors by half.

The entanglement entropy is evaluated by applying the HOTRG algorithm to the reduced density matrix $\rho_A$.
Let $T^{(0)}$ be an initial tensor representing $\rho_A$.
The algorithm is applied alternately to the spatial and temporal directions, and a single set consists of two renormalizations (one renormalization in each direction).
Applying $k$ sets of renormalizations, we obtain the renormalized tensor $T^{(k)}$.
Let $U^{(k-1)}$ denote the isometry matrix used for the renormalization in the spatial direction of $k$-th renormalization set.
We use these isometry matrices in our method later,
but the isometry matrices of the temporal direction are not needed for any purpose other than to obtain $T^{(k)}$.

\begin{figure}[ht]
	\centering
	\begin{minipage}[b]{0.48\linewidth}
		\centering
		\includegraphics[keepaspectratio, scale=0.51]{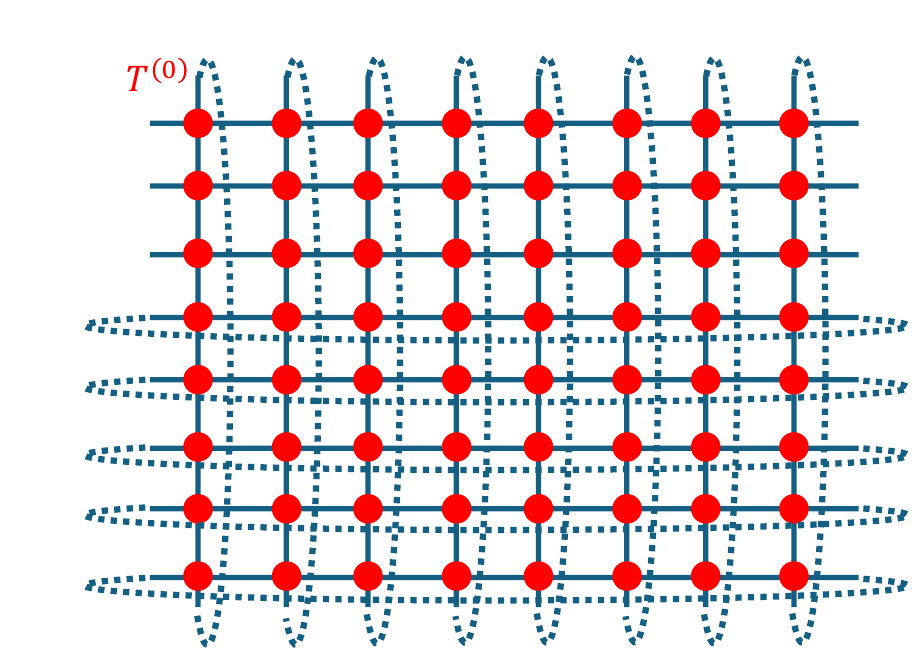}
		\subcaption{Tensor network representation of ${\rho}_A$. }
		\label{fig.RDM_HOTRG_0}
	\end{minipage}
	\begin{minipage}[b]{0.48\linewidth}
		\centering
		\includegraphics[keepaspectratio, scale=0.55]{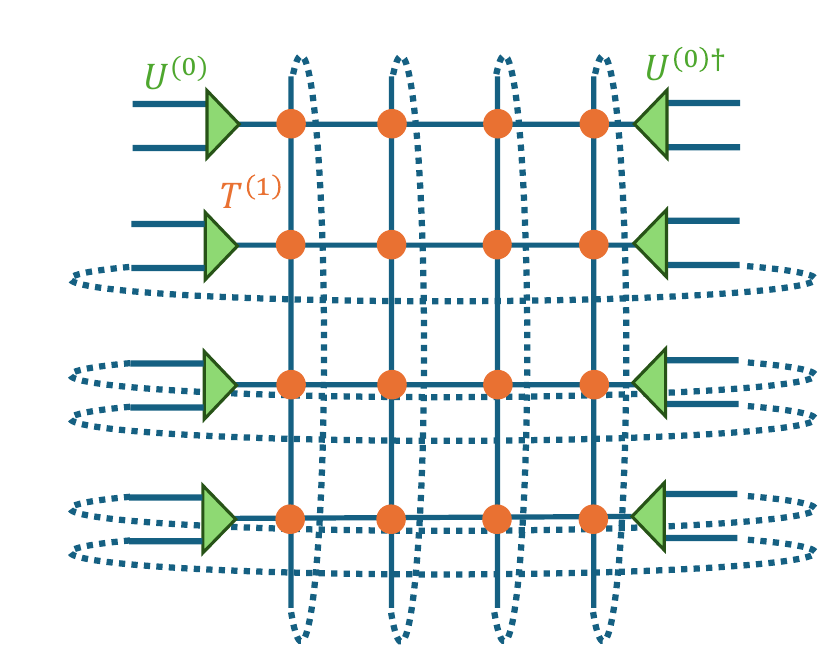}
		\subcaption{$\rho_A$ after a set of renormalization ($k=1$).}
		\label{fig.RDM_HOTRG_1}
	\end{minipage}
	\vspace{20pt}\\
	\begin{minipage}[b]{0.48\linewidth}
		\centering
		\includegraphics[keepaspectratio, scale=0.6]{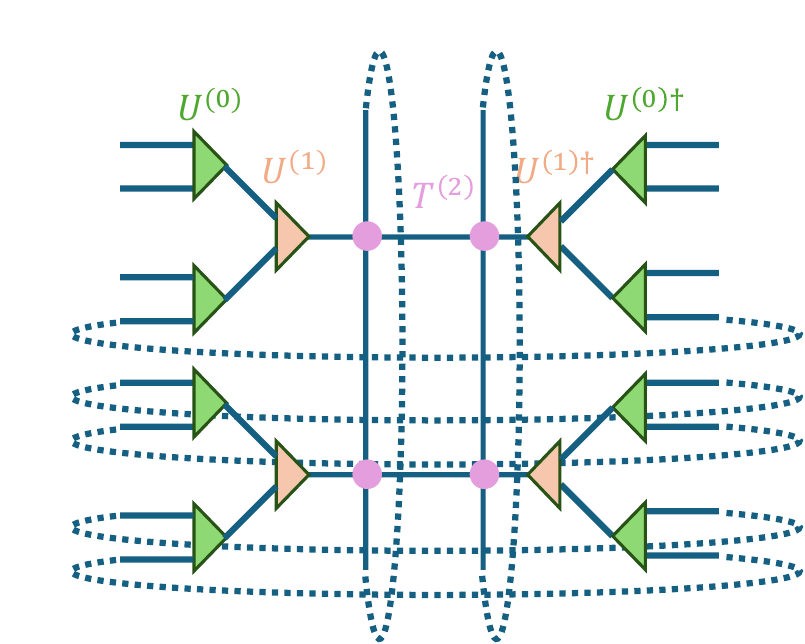}
		\subcaption{$\rho_A$ after two sets of renormalization ($k=2$).}
		\label{fig.RDM_HOTRG_2}
	\end{minipage}
	\begin{minipage}[b]{0.48\linewidth}
		\centering
		\includegraphics[keepaspectratio, scale=0.53]{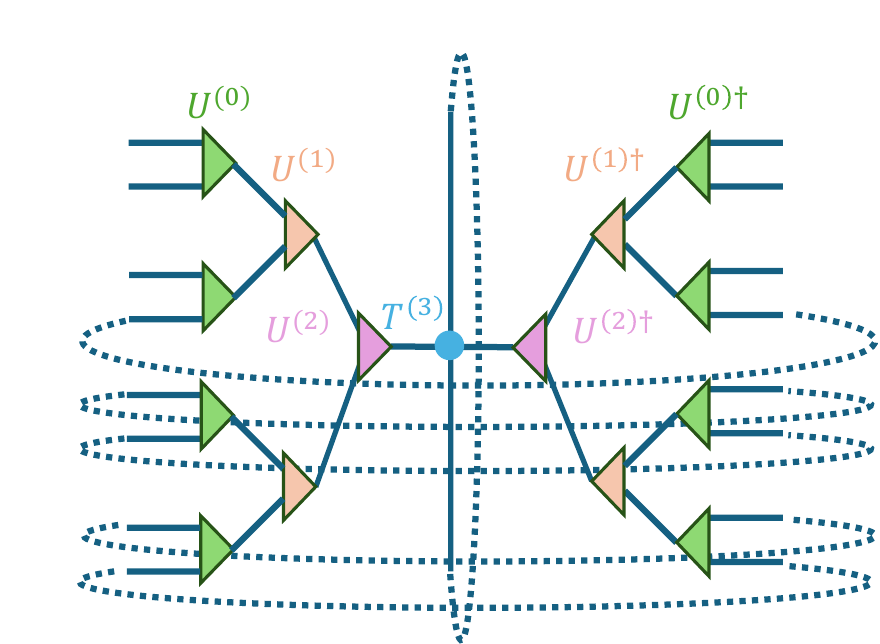}
		\subcaption{$\rho_A$ after three sets of renormalization ($k=3$).}
		\label{fig.RDM_HOTRG_3}
	\end{minipage}
	\caption{Coarse-graining procedure for ${\rho}_A$.}
	\label{fig.HOTRG_matomete}
\end{figure}
To illustrate our method, we first consider the case of $L=N=8$ and $\ell=3$ as shown in Fig.~\ref{fig.RDM_HOTRG_0},
where six external lines remain open, and $\rho_A$ is a $D^3 \times D^3$ matrix.
Figure~\ref{fig.RDM_HOTRG_1}, \ref{fig.RDM_HOTRG_2} and \ref{fig.RDM_HOTRG_3}
denote the approximations of $\rho_A$ after one, two, and three sets of renormalizations, respectively.

\begin{figure}[p]
	\centering
	\begin{minipage}[b]{0.96\linewidth}
		\centering
		\includegraphics[keepaspectratio, scale=0.6]{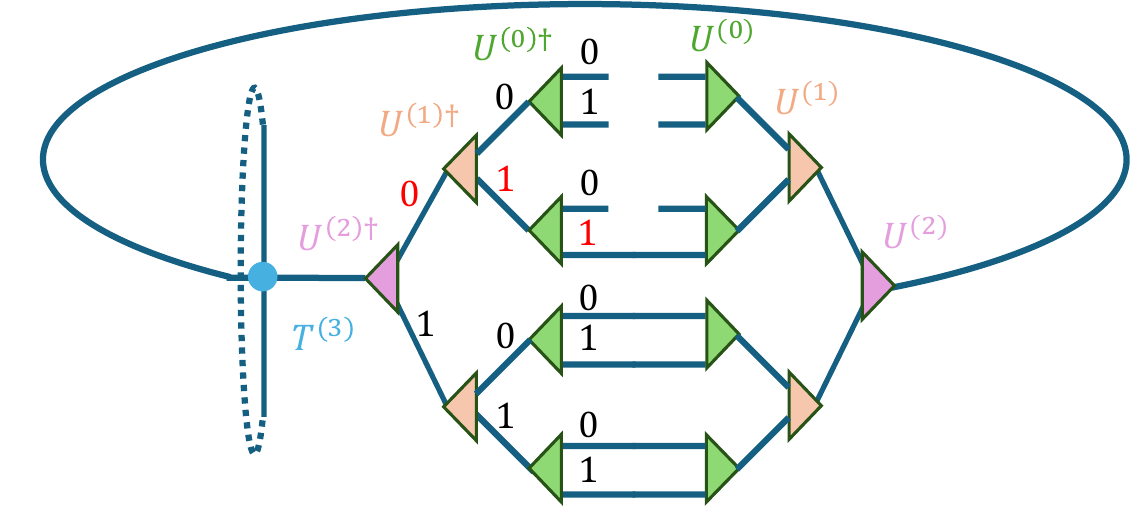}
		\subcaption{Tensor network of ${\rho}'_A$ that is geometrically equivalent to Fig. \ref{fig.RDM_HOTRG_3}.
		For later convenience, each line emerging from the isometry matrices $U^{(i)\dagger}$ is labeled as 0 or 1 for upper and lower lines, respectively.}
		\label{fig.geometrically_equivalent_0}
	\end{minipage}
	\vspace{20pt}\\
	\begin{minipage}[b]{0.96\linewidth}
		\centering
		\includegraphics[keepaspectratio, scale=0.6]{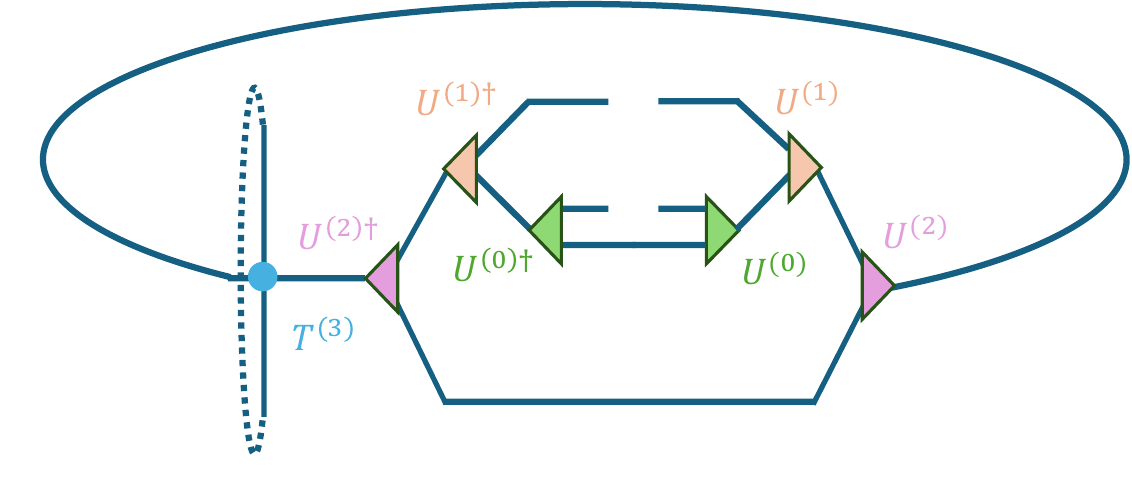}
		\subcaption{Tensor network of $\tilde{\rho}_A$ obtained by dropping irrelevant isometry matrices and using $U^{(i)\dagger} U^{(i)}=I$ in Fig.~\ref{fig.geometrically_equivalent_0}.}
		\label{fig.geometrically_equivalent_1}
	\end{minipage}
	\vspace{20pt}\\
	\begin{minipage}[b]{0.96\linewidth}
		\centering
		\includegraphics[keepaspectratio, scale=0.6]{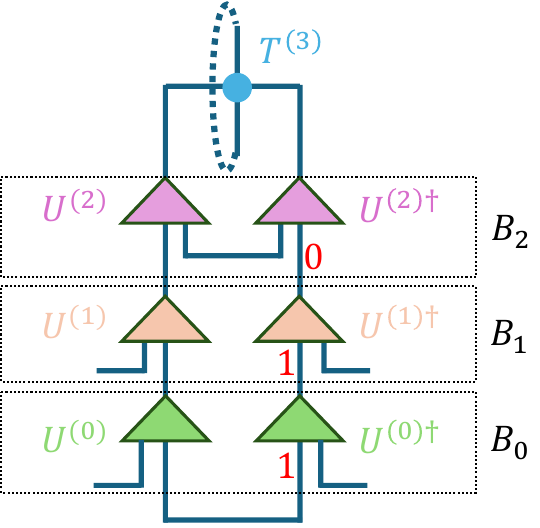}
		\subcaption{Tensor network of $\tilde{\rho}_A$ geometrically equivalent to Fig.~\ref{fig.geometrically_equivalent_1}.
		The binary representation of $\ell$ determines the contractions of isometry matrices $U^{(k)}$ and $U^{(k)\dagger}$ in each dotted box $B_k\ (k=0,1,2)$.}
		\label{fig.geometrically_equivalent_2}
	\end{minipage}
	\vspace{20pt}\\
	\caption{${\rho}'_A$ and $\tilde{\rho}_A$.}
	\label{fig.geometrically_equivalent}
\end{figure}

The renormalized $\rho_A$ in Fig.~\ref{fig.RDM_HOTRG_3} consists of $U^{(0)}, U^{(1)}, U^{(2)}$  and $T^{(3)}$, and we denote it as $\rho'_A$.
The isometry matrices $U^{(i)}$ and $U^{(i)\dagger} \ (i=0,1,2)$ always come in pairs.
Fig.~\ref{fig.geometrically_equivalent_0} is geometrically equivalent to Fig.~\ref{fig.RDM_HOTRG_3}, and can be simplified to Fig.~\ref{fig.geometrically_equivalent_1}.
Loops made of $U^{(i)}$ and $U^{(i)\dagger}$ in Fig.~\ref{fig.geometrically_equivalent_0} become identity matrix since $U^{(i)\dagger} U^{(i)}=I$.
In addition, pairs of $U^{(i)}$ and $U^{(i)\dagger}$ with two open indices can be dropped
because they do not contribute to the entanglement entropy.
This can be seen as follows: letting $\tilde{\rho}_A$ denote the remainder of the network,
$\mathrm{Tr} ({\rho'}_A \log ({\rho'}_A))=\mathrm{Tr}( U^{(i)} \tilde{\rho}_A U^{\dagger(i)} \log (U^{(i)} \tilde{\rho}_A U^{\dagger(i)}))
=\mathrm{Tr} (\tilde{\rho}_A \log (\tilde{\rho}_A))$, where we have used the property $U^{(i)\dagger} U^{(i)} = I$.
Thus, we find that Fig.~\ref{fig.geometrically_equivalent_0} becomes Fig.~\ref{fig.geometrically_equivalent_1} without further approximations.
Figure~\ref{fig.geometrically_equivalent_2} is an equivalent representation of Fig.~\ref{fig.geometrically_equivalent_1}.

Consequently, $S_A$ is evaluated from $\tilde{\rho}_A$ shown in Fig.~\ref{fig.geometrically_equivalent_2} as $S_A\simeq -\mathrm{Tr}( \tilde{\rho}_A \log (\tilde{\rho}_A))$.
The trimmed network $\tilde{\rho}_A$ has a simpler structure than the original network $\rho_A$ shown in Fig.~\ref{fig.RDM_HOTRG_0} and renormalized one in Fig.~\ref{fig.RDM_HOTRG_3}.
Compared to Fig.~\ref{fig.RDM_HOTRG_3}, the number of isometry matrices is reduced from fourteen to six,
and the size of the reduced density matrix is reduced from $D^3 \times D^3$ to $D^2 \times D^2$.

\begin{figure}[ht]
	\centering
	\begin{minipage}[b]{0.48\linewidth}
		\centering
		\includegraphics[keepaspectratio, scale=0.7]{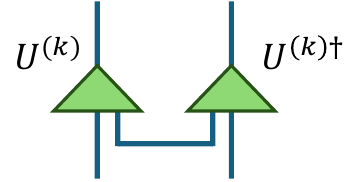}
		\subcaption{Case of $a_k = 0$.}
		\label{fig.ISO_CONT}
	\end{minipage}
	\centering
	\begin{minipage}[b]{0.48\linewidth}
		\centering
		\includegraphics[keepaspectratio, scale=0.7]{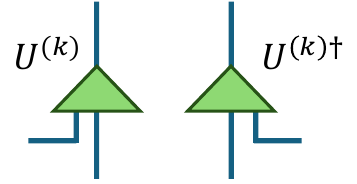}
		\subcaption{Case of $a_k = 1$.}
		\label{fig.ISO_PROD}
	\end{minipage}
	\caption{Inside of each dotted box $B_k$ defined in Fig.~\ref{fig.geometrically_equivalent_2}, the isometry matrices $U^{(k)}$ and $U^{(k)\dagger}$ are contracted as illustrated in Fig.~\ref{fig.ISO_CONT} if $a_k=0$,
		and as illustrated in Fig.~\ref{fig.ISO_PROD} if $a_k=1$.
		In both cases, the line towards the top is contracted with the line towards the bottom emerging from $U^{(k+1)}$ and $U^{(k+1)\dagger}$,
		or $T^{(3)}$ (or the matrix $C$, which we define later).
		In Fig.~\ref{fig.ISO_PROD}, the lines towards the left and right remain open and become tensor indices of $\tilde{\rho}_A$.}
		\label{fig.B_k}
\end{figure}

The form of the final expression (Fig.~\ref{fig.geometrically_equivalent_2}) is closely related to the binary representation of $\ell$.
In this case, we have $\ell=(a_2 a_1 a_0)_2=(011)_2$.
Figure~\ref{fig.geometrically_equivalent_2} has three dotted boxes $B_k \ (k=0,1,2)$,
and the internal structure of each box is given by Fig.~\ref{fig.ISO_CONT} or Fig.~\ref{fig.ISO_PROD}.
The case of $a_k=0$ corresponds to Fig.~\ref{fig.ISO_CONT}, and the case of $a_k=1$ corresponds to Fig.~\ref{fig.ISO_PROD}.
Thus, we can see that the ordering of three boxes in Fig.~\ref{fig.geometrically_equivalent_2} coincides with the binary representation of $\ell=(011)_2$.
The reason why this structure occurs can be understood as follows: In Fig.~\ref{fig.geometrically_equivalent_0},
we focus on two lines extending to the right from the same isometry matrix, and
label the upper line as 0 and the lower line as 1.
Let us take a look at closed lines connecting $U^{(0)\dagger}$ and $U^{(0)}$.
The topmost closed line starting from $T^{(3)}$ is the path 011 denoted by red numbers.
The path 011, which is the binary representation of $\ell$, also appears as the vertical line in Fig.~\ref{fig.geometrically_equivalent_2}.
We thus find that the binary representation of $\ell$ corresponds to
the internal form of three boxes.

\begin{figure}[ht]
		\centering
		\begin{minipage}[b]{0.48\linewidth}
			\centering
			\includegraphics[keepaspectratio, scale=0.7]{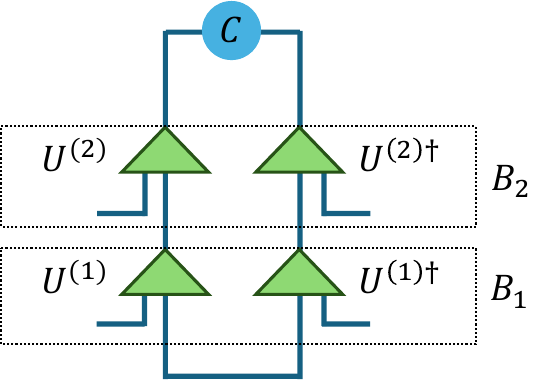}
			\subcaption{$\ell=2$. }
			\label{fig.general_rho_l2}
		\end{minipage}
		\begin{minipage}[b]{0.48\linewidth}
			\centering
			\includegraphics[keepaspectratio, scale=0.7]{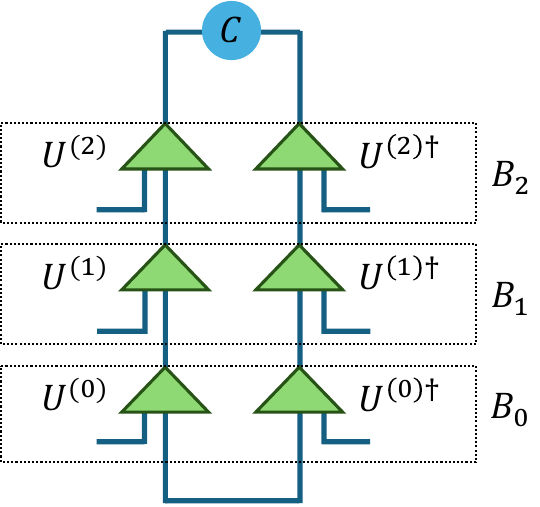}
			\subcaption{$\ell=7$. }
			\label{fig.general_rho_l7}
		\end{minipage}
		\caption{Trimmed network $\tilde{\rho}_A$ for $n=3$, $\ell=2$ and $\ell=7$.}
		\label{fig.rho_example}
\end{figure}

The number of boxes also depends on $\ell$.
We show $\tilde{\rho}_A$ for $\ell=2$ and 7 in Fig.~\ref{fig.general_rho_l2} and \ref{fig.general_rho_l7}, respectively.
This number is actually given as $3-r$, where $r$ is the position of the rightmost set bit (1-bit) of $\ell$.
The matrix $\tilde{\rho}_A$ has three boxes when $\ell=7=(111)_2$ since $r=0$,
while it has two boxes when $\ell = 2 = (010)_2$ since $r=1$.

\begin{figure}[ht]
		\centering
		\begin{minipage}[b]{0.48\linewidth}
			\centering
			\includegraphics[keepaspectratio, scale=0.55]{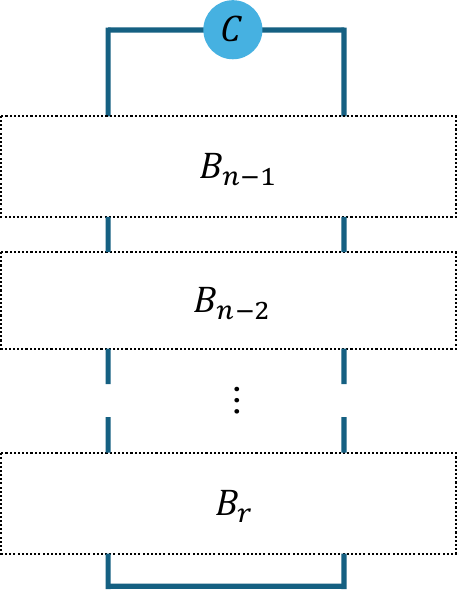}
			\caption{Trimmed network $\tilde{\rho}_A$. }
			\label{fig.general_rho}
		\end{minipage}
		\begin{minipage}[b]{0.48\linewidth}
			\centering
			\includegraphics[keepaspectratio, scale=0.7]{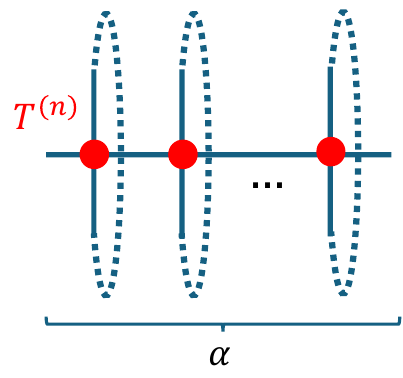}
			\caption{Matrix $C$.}
			\label{fig.Core_Tn}
		\end{minipage}
\end{figure}

We generalize the result above to $L=2^n$ and $N=\alpha \cdot 2^n \ (\alpha \in \mathbb{N})$,
and any subsystem size $\ell =1,2,\dots, L-1$.
As a result, the entanglement entropy is given as $S_A \simeq - \mathrm{Tr} (\tilde{\rho}_A \log (\tilde{\rho}_A))$,
where $\tilde{\rho}_A$ is defined by Fig.~\ref{fig.general_rho}.
A matrix $C$ is given by Fig.~\ref{fig.Core_Tn}, and $r$ is defined as the position of the rightmost set bit of $\ell$ in binary form.
The internal form of each dotted box $B_k \ (k=r, r+1, \dots, n-1)$ is determined by Fig.~\ref{fig.B_k} according to $a_k$ of
\begin{align}
	\ell =& \sum_{i=0}^{n-1} 2^i a_i = (a_{n-1} a_{n-2} \cdots a_1 a_0)_2,
	\label{eq.binary_l}
\end{align}
where $a_i = 0, 1$.

The original $\rho_A$ is a $D^\ell \times D^\ell$ matrix, and the number of isometry matrices constituting $\rho'_A$ is $O(L)$.
Our method reduces the size of the matrix to $D^h \times D^h$, where $h=\sum_i a_i$ is the Hamming weight (the number of set bits) of $\ell$ in the binary representation.
The number of isometry matrices in $\tilde{\rho}_A$ is also reduced from $O(L)$ to $O(\log L)$.

We finally mention the computational cost and a modification of our method.
Since we use HOTRG, the cost for renormalization is the same as the standard HOTRG algorithm.
Additional computations are required to obtain $S_A$.
Evaluating $\tilde{\rho}_A$ costs $O(D^{2h+2})$, while diagonalizing $\tilde{\rho}_A$ to calculate $S_A$ costs $O(D^{3h})$.
To improve accuracy, we can skip the final coarse-graining set.
In that case, the matrix $C$ and box $B_{n-1}$ of Fig.~\ref{fig.general_rho} are replaced with a matrix $C'$ that is a product of $2\alpha$ tensors $T^{(n-1)}$.
The computational cost of the product is $O(D^{6})$.
Other HOTRG-like algorithms, which renormalize two vertically aligned tensors into a single tensor using an isometry, such as Triad TRG \cite{Kadoh:2019kqk} and MDTRG \cite{Nakayama:2023ytr}, can also be employed to reduce the computational cost.

\clearpage
\section{Numerical test}
\label{sec:result}
The transverse field Ising model in $d=1$ is equivalent to (1+1)-dimensional classical model with anisotropic coupling \cite{suzukiRelationshipDDimensionalQuantal1976}.
In previous studies \cite{PhysRevB.89.075116, Bulgarelli:2023ofi},
the isotropic ising model is often used to study the scaling property of EE.
In this section, we test our method in the isotropic classical ising model.

\subsection{Entanglement entropy}
We consider a ``density matrix'' $\rho$ defined as a two-dimensional tensor network Fig.~\ref{fig.time_evolution_tensor_network_finite_time} of
\begin{align}
	T_{ijkl}
	=& C \delta_{\mathrm{mod}(i+j+k+l, 2),0} \cosh^2 \beta
		\left( \tanh \beta \right)^{(i+j+k+l)/2},
	\label{eq.tensor_representation}
\end{align}
with the inverse temperature $\beta=J/T$, and a constant $C$ is chosen so that $\mathrm{Tr}(\rho)=1$.
This tensor is obtained from the (1+1)-dimensional isotropic classical Ising model $H=-J\sum_{<a,b>}s_a s_b$, where we set $J=1$ for simplicity.
The tensor network of Fig.~\ref{fig.time_evolution_tensor_network_finite_time} is closed in the vertical direction (spatial direction), and the horizontal direction corresponds to the imaginary time direction.
Let $L$ and $N=\alpha L \ (\alpha \gg 1)$ be the sizes of the spatial and temporal directions, respectively.
External lines of $\rho$ shown in Fig.~\ref{fig.time_evolution_tensor_network_finite_time} correspond to the local degrees of freedom of the quantum system.
Partial trace over a subsystem is performed by contracting the corresponding indices.

We consider a single interval of length $\ell$ as a subsystem $A$.
The entanglement entropy $S_A$ is defined as $S_A = - \mathrm{Tr}(\rho_A \log(\rho_A))$, where $\rho_A=\mathrm{Tr}_{\bar{A}} (\rho)$ is the reduced density matrix.
Figure~\ref{fig.reduced_density_matrix_tensor_network} shows the tensor network representation of $\rho_A$,
which is obtained by partial trace of Fig.~\ref{fig.time_evolution_tensor_network_finite_time} with respect to $\bar{A}$.
In this case, $S_A$ is a function of $L$ and $\ell$, and the CFT mapped into a cylinder of spatial length $L$ predicts the following analytical expression for $S_A(L, \ell)$:
\begin{align}
	S_A(L, \ell) = \frac{c}{3}\log \left( L \sin \left( \frac{\ell}{L} \pi \right) \right) + k_1 \ ,	\label{EE_cylinder}
\end{align}
where $c$ is the central charge of the theory, and $k_1$ is a non-universal constant\cite{Calabrese:2004eu}.

The entanglement entropy $S_A(L, \ell)$ is computed by our method presented in Sec.~\ref{sec:theory}.
We use the standard HOTRG algorithm \cite{Xie:2012mjn} for renormalization.
The size $L$ is set to $2^n$, and $\ell$ is set to $2^m$ or $2^m + 2^q$, where $n,m$ and $q$ are non-negative integers satisfying $q<m<n$.
The size of temporal direction is $N=\alpha L$, where the parameter $\alpha$ should be sufficiently large in order to compare the numerical results with the analytical prediction of CFT \eqref{EE_cylinder}.
Since we employ the HOTRG algorithm, the computational cost of each renormalization is $O(D^7)$, where $D$ is the bond dimension of the tensor $T_{ijkl}$.
In addition, computing $S_A$ costs $O(D^{\max (2h+2, 3h)})$, where $h$ is the Hamming weight of $\ell$ expressed in the binary representation.
Since $\ell=2^m$ or $2^m + 2^q$, we have $h=1$ or 2, and thus the overall computational cost of evaluating $S_A$ remains $O(D^7)$.

\subsection{Results}
The entanglement entropy is evaluated at $T_c = {2}/{\log(1+\sqrt{2})}$.
First, we present $S_A(L, \ell)$ calculated for various $\ell$ at fixed $L$, and then we show the results varying $L$ for fixed $x = \ell/L$.

\begin{figure}[H]
	\begin{center}
		\includegraphics[width=11cm]{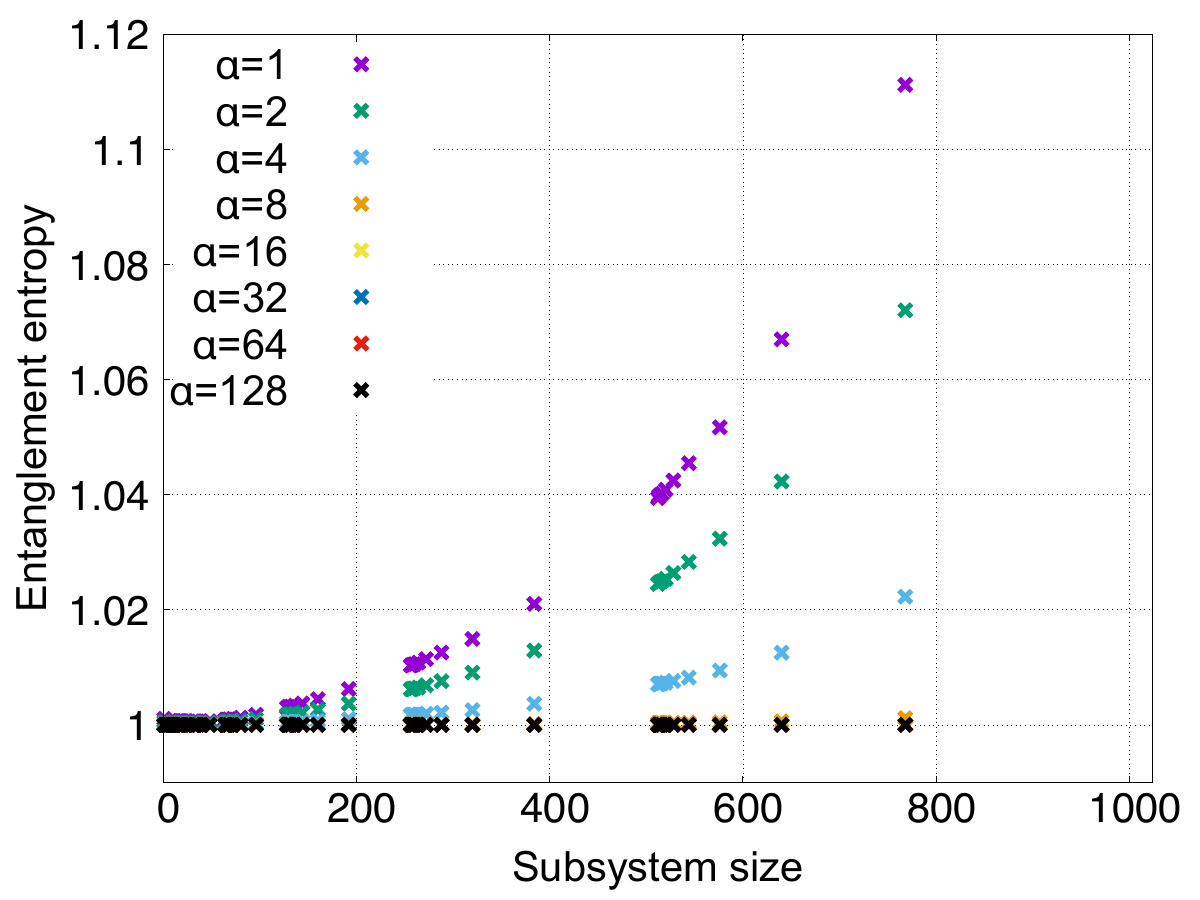}
	\end{center}
	\caption{Entanglement entropy at $T=T_c$ as a function of $\ell$ for several values of $\alpha=1,2,4, \dots, 128$.
	The bond dimension and the lattice size are fixed to $D=64$ and $L=1024$ respectively.
	All results are normalized by the values at $\alpha=1024$.}
	\label{EE_alpha}
\end{figure}

We consider a case of $L=1024$.
Fig.~\ref{EE_alpha} shows the $\alpha$ dependence of the entanglement entropy $S_A(L,\ell)$ at $D=64$.
For $\alpha\gtrsim16$, the values of $S_A(L, \ell)$ show convergence with differences being smaller than $10^{-10}$.
In the following, we thus fix $\alpha=16$ which is large enough to obtain the converged results.

\begin{figure}[p]
\begin{center}
\includegraphics[width=11cm]{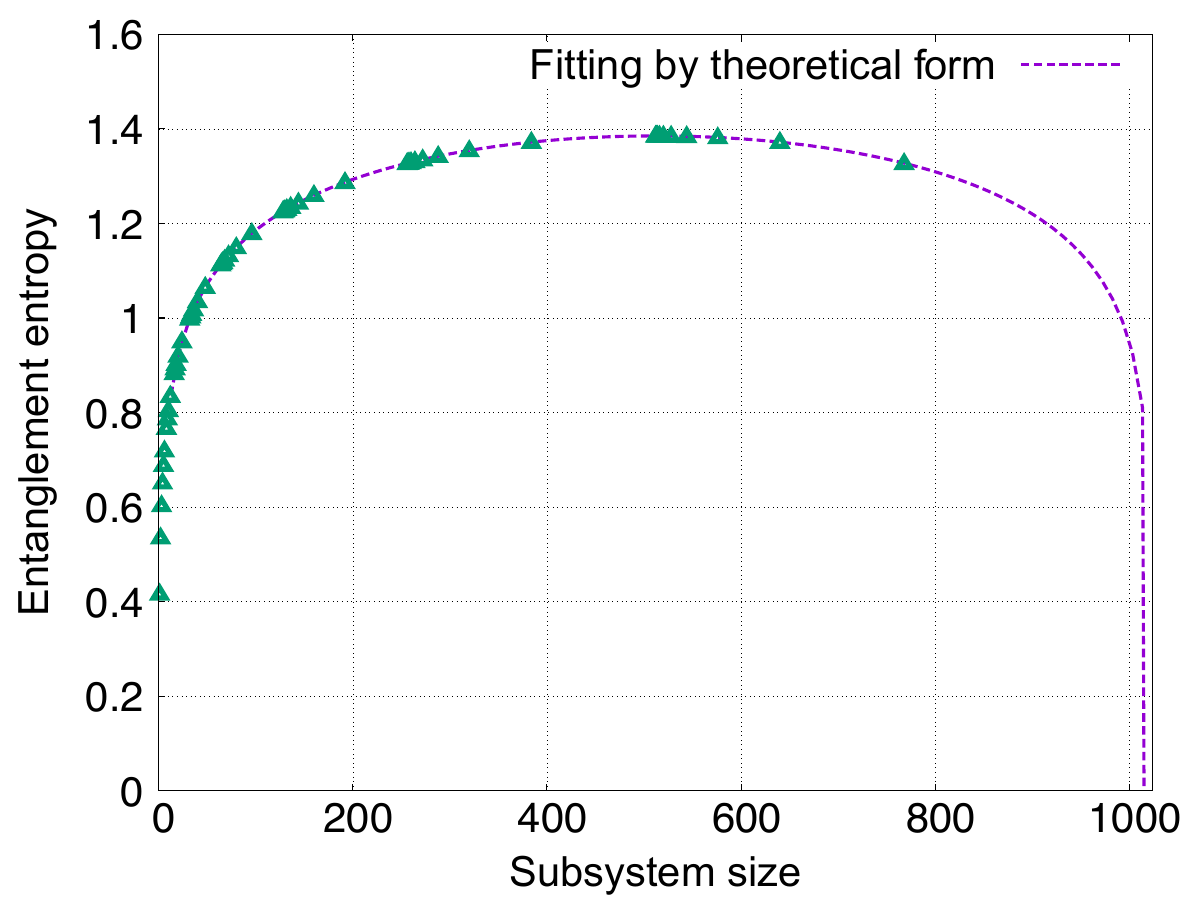}
\end{center}
\caption{Entanglement entropy at $T=T_c$ as a function of $\ell$ at $D=96$ and $\alpha=16$.}
\label{arealaw}
\begin{center}
\includegraphics[width=11cm]{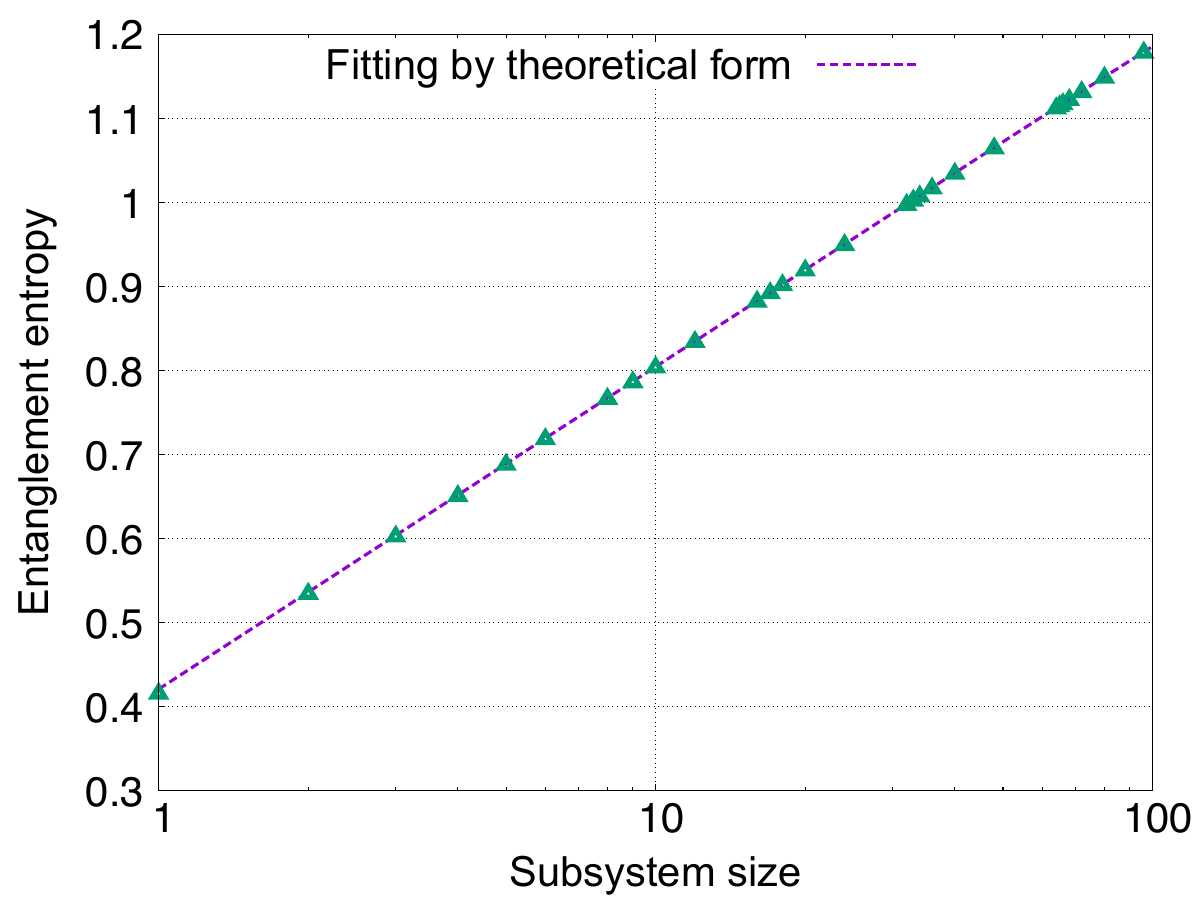}
\end{center}
\caption{Zoom of Fig.~\ref{arealaw} at $T=T_c$ for small $\ell$.
}
\label{arealaw_small}
\end{figure}

Figure~\ref{arealaw} shows the dependence on $\ell$ of $S_A(L, \ell)$ and the fitting result with the theoretical functional form \eqref{EE_cylinder}, where $c$ and $k_1$ are fit parameters.
The theoretical form describes the data well in the whole region and
the symmetric property ($A\leftrightarrow B$) is clearly observed.
Figure \ref{arealaw_small} is a zoom of Fig.~\ref{arealaw} around small $\ell$ ($\ll L$) region, where the horizontal axis is the logarithmic scale.
From the figure, $\log(L\sin(\ell/L \pi))\approx \log \ell$ behavior is clearly seen, and
the central charge can be extracted from its slope.

The fitting range of $\ell$ should be determined to obtain the central charge.
For that purpose, we first compute an effective central charge as:
\begin{align}
	c(L, \ell) = 3 \frac{S_A(L, \ell') - S_A(L, \ell) }{\log \left( \sin \left( \frac{\ell'\pi}{L} \right) \right) - \log \left( \sin \left( \frac{\ell\pi}{L} \right) \right)}
\label{eqn:cl}
\end{align}
with $\ell=2^m+q$ and $\ell'=2^{m+1}+q$, where $1 < m < 9$, $q=0, 1, 2, 2^2, 2^3, \ldots , 2^6 <2^m$.
\begin{figure}[ht]
	\begin{center}
		\includegraphics[width=11cm]{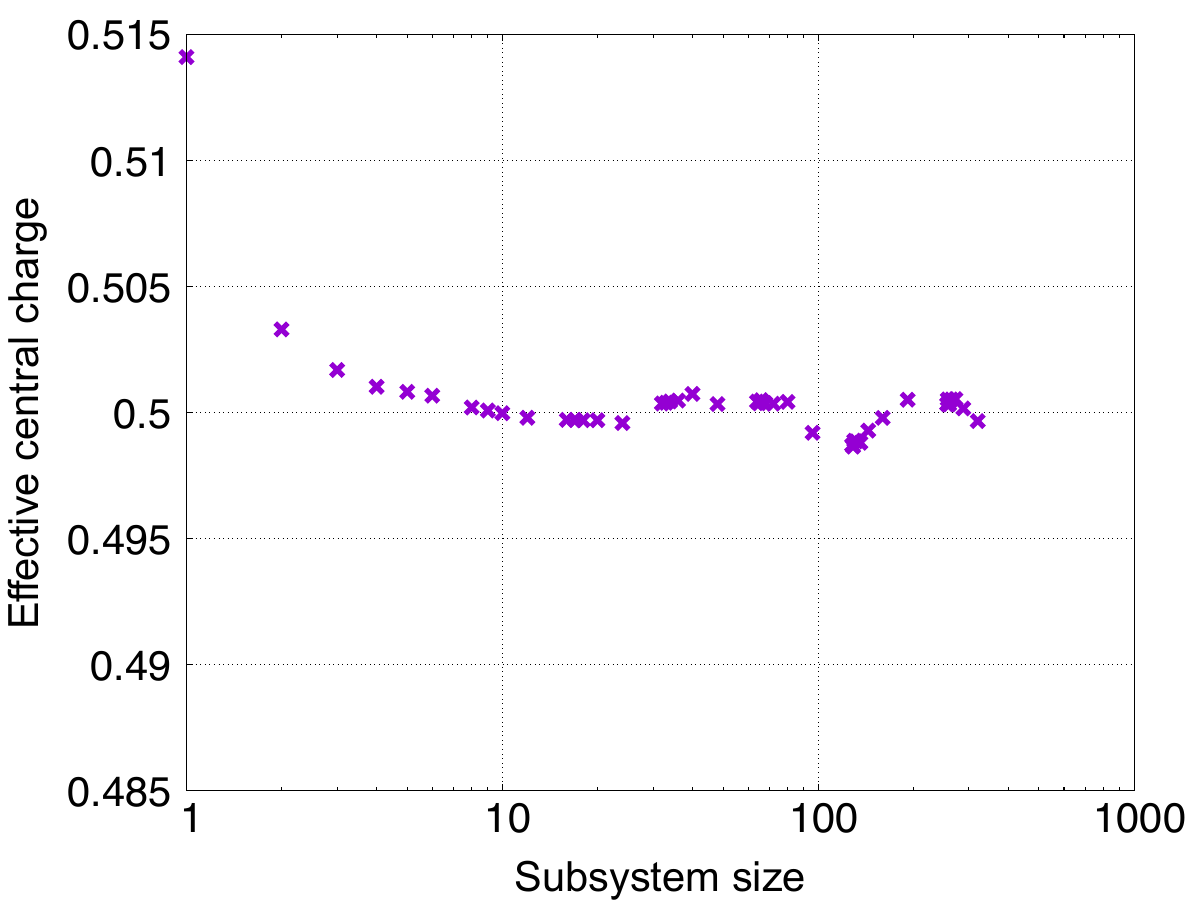}
	\end{center}
	\caption{Effective central charge $c(\ell)$ computed by (\ref{eqn:cl}) at $T=T_c$.}
	\label{l_range}
\end{figure}
The numerical result of $c(L, \ell)$ is shown in Fig.~\ref{l_range}.
For small $\ell$, $c(L, \ell)$ is inconsistent with the expected constant behavior.
We choose $7 \leq \ell \leq 768$ for the fit range and use the fitting form (\ref{EE_cylinder}) with two fit parameters $c$ and $k_1$.
The obtained values for $D=96$ are
\begin{align}
	c=0.49997(8), \hspace{20pt} k_1=0.2300(2).	\label{eq.result_D96}
\end{align}
To estimate the error of $c$, we solve (\ref{EE_cylinder}) with respect to $c$ for each $\ell$ using numerical value of $S_A(L,\ell)$, where $k_1$ is fixed to the central value obtained from the fitting.
The error is given by the maximal difference between the solved central charge and the central value in \eqref{eq.result_D96}.
The error of $k_1$ is also estimated in the same way.
The same analysis can be repeated for other bond dimension $D=64$ and $80$, and the results are summarized in Table \ref{CCforDcut}.
\begin{table}[ht]
\begin{center}
\caption{$D$-dependence of the central charge extracted from the entanglement entropy.}
\label{CCforDcut}
\begin{tabular}{cl}
	\hline	\hline
	$D$ & central charge	\\	\hline
	$64$	&0.4998(2)	\\
	$80$	&0.4999(1)	\\
	$96$	&0.49997(8)	\\	\hline\hline
\end{tabular}
\end{center}
\end{table}

As a second analysis, let us consider the case of fixed $x \equiv \ell/L$.
For fixed $x$, we have
\begin{align}
	S_{A,x}(L) \equiv S_A(L, xL)
	&= \frac{c}{3} \log L + k'_1(x) \ ,	\label{EE_xfix}
\end{align}
with
\begin{align}
	k'_1(x)
	= k_1 + \frac{c}{3} \log \left( {\sin x \pi } \right) \ .
\end{align}
The effective central charge can be extracted from the difference of entanglement entropy at $L$ and $2L$:
\begin{align}
	c_x(L) = \frac{3}{\ln 2} \left( S_{A,x}(2L) - S_{A,x}(L) \right)	\ .	\label{CC_from_diff}
\end{align}

\begin{figure}[p]
	\begin{center}
		\includegraphics[width=11cm]{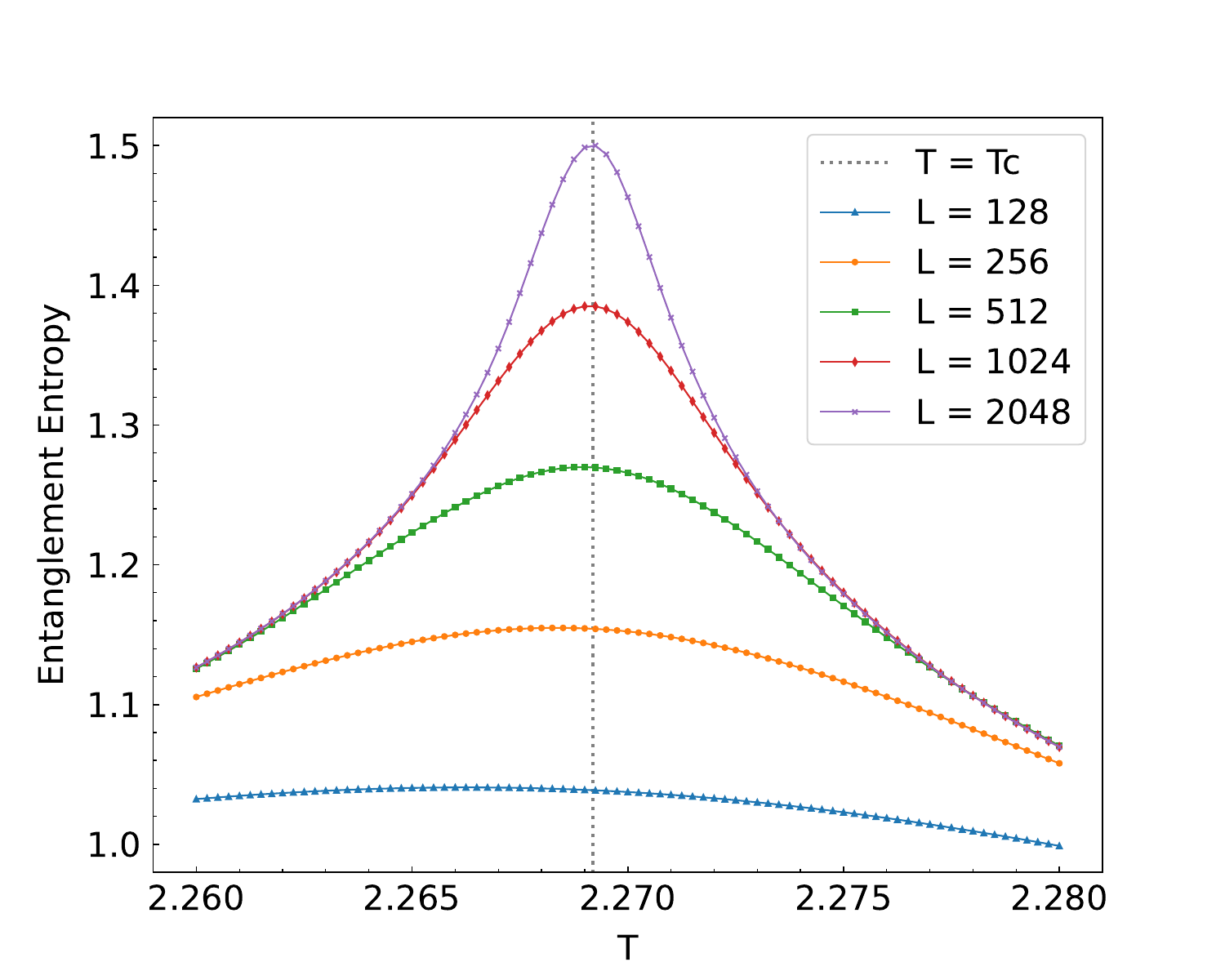}
		\caption{Temperature dependence of entanglement entropy around the critical point
				at $\alpha = 16$ and $D = 96$.
				The ratio $x=\ell/L=1/2$ is fixed and the total size is varied in the range $L=128-2048$.
				The dotted grey line shows the location of the critical temperature $T_c$.}
		\label{fig:EE}
	\end{center}
	\begin{center}
		\includegraphics[width=11cm]{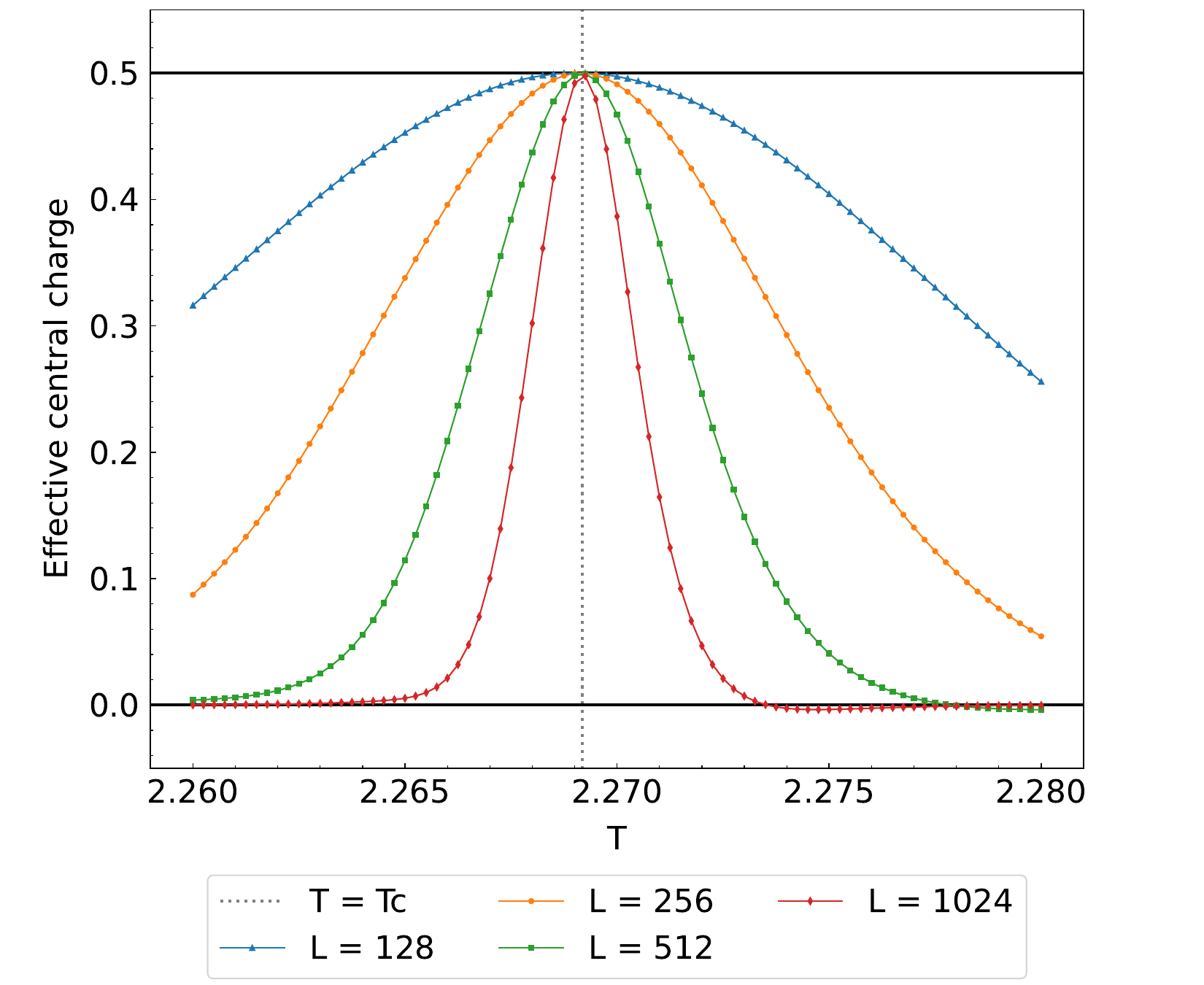}
		\caption{Temperature dependence of effective central charge defined in \eqref{CC_from_diff} near critical point. The parameters $T$, $\alpha$, $D$ and $x$ are the same as those in Fig.~\ref{fig:EE}.}
		\label{fig:CC}
	\end{center}
\end{figure}

Figures~\ref{fig:EE} and \ref{fig:CC} show the temperature dependence of the entanglement entropy and the effective central charge \eqref{CC_from_diff} for $L=2^7,2^8,\dots, 2^{11}$ with fixed $x=\ell/L=1/2$ and $D=96$.
It can be seen from Figs.~\ref{fig:EE} and \ref{fig:CC} that peaks appear at the critical point $T_c$ for large volume $L$. At $T_c$, the effective central charge is close to $c=0.5$.

\begin{figure}[p]
	\begin{center}
		\includegraphics[width=11cm]{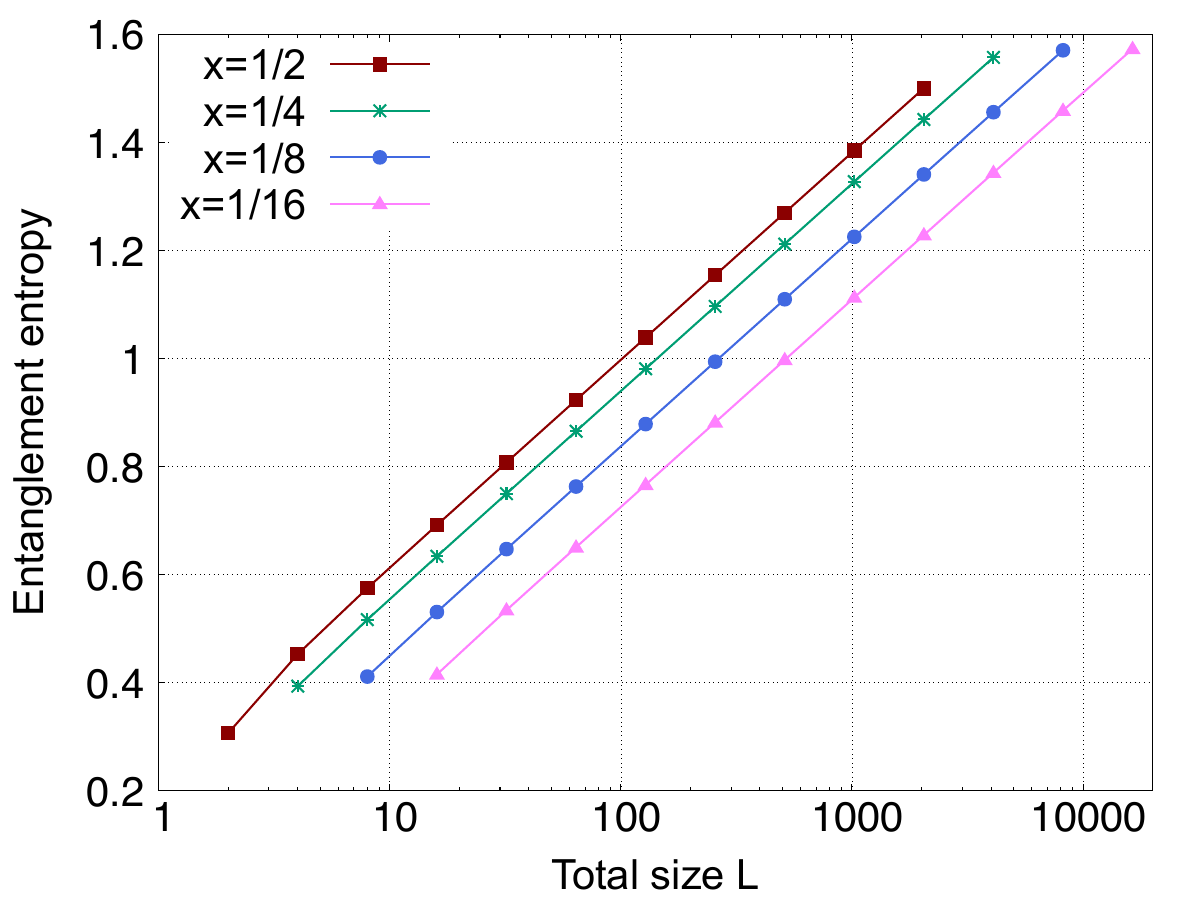}
	\end{center}
	\caption{$L$ dependence of the entanglement entropy at $T=T_c$ for various values of the ratio $x = 1/2, 1/4, 1/8$, and $1/16$
	at $D = 96$ and $\alpha = 16$.}
	\label{x_depend}
	\begin{center}
		\includegraphics[width=11cm]{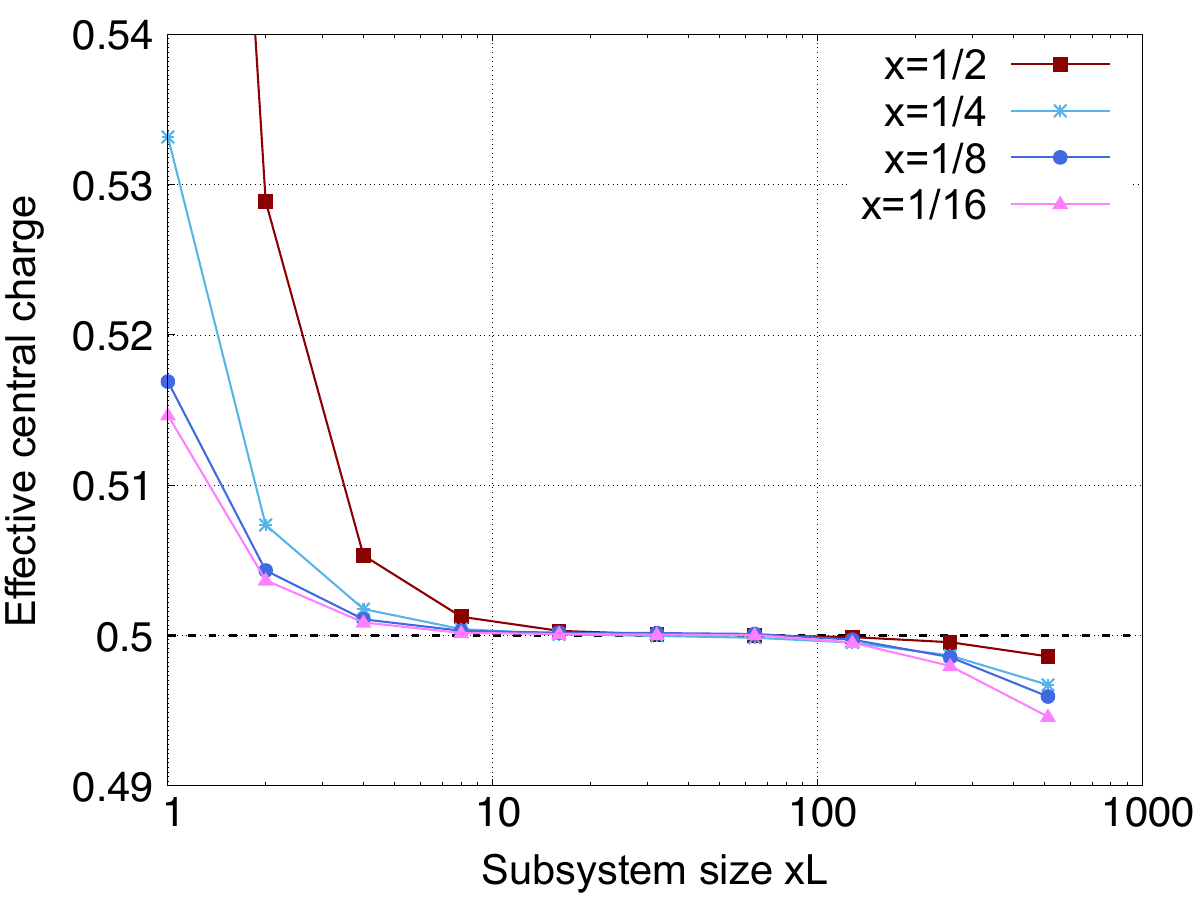}
	\end{center}
	\caption{Effective central charge $c_x(L)$ in (\ref{CC_from_diff}) at $T=T_c$ as a function of $L$ with $x = 1/2, 1/4, 1/8$ and $1/16$ at $D = 96$ and $\alpha = 16$.}
	\label{CC_x}
\end{figure}

Figure~\ref{x_depend} shows the $L$ dependence of $S_A(L,x)$ for $x = 1/2, \ 1/4, \ 1/8$, and $1/16$ at $D = 96$ and $T=T_c$.
We can also extract the central charge using (\ref{EE_xfix}) as a fitting function with fitting parameters $c$ and $k'_1(x)$.
Similarly to the first analysis, to determine the fit range, we compute an effective central charge using \eqref{CC_from_diff}, and results are presented in Fig.~\ref{CC_x}.

The effective central charges $c_x(L)$ exhibit a plateau in the range of $16 \leq xL \leq 128$, and the plateau value is close to $c=0.5$.
We thus choose the plateau region as a fit range and carry out the fitting with the functional form of (\ref{EE_xfix}).
The error of $c$ is estimated in the same way as in the first analysis.
The resulting central charge shown in Table \ref{ccharge_xfix} is consistent with the expected value of $c=0.5$.
\begin{table}[H]
\begin{center}
\caption{Central charge extracted from the entanglement entropy with fixed $x=\ell/L$ at $D=96$ and $\alpha=16$.}
\label{ccharge_xfix}
\begin{tabular}{ccc}
	\hline	\hline
	$x$ & central charge	\\	\hline
	$1/2$	&0.5001(2)	\\
	$1/4$	&0.5000(4)	\\
	$1/8$	&0.5001(4)	\\
	$1/16$	&0.5001(5)	\\	\hline\hline
\end{tabular}
\end{center}
\end{table}

Throughout the first and third analyses, our method reproduced the expected logarithmic behavior of the entanglement entropy \eqref{EE_cylinder} and the central charge $c=0.5$.
We thus conclude that our method works well.

\section{Summary and outlook}
\label{sec:summary}
In this paper, we have presented a tensor renormalization group method
to compute the entanglement entropy for an arbitrary subsystem size.
We have considered one-dimensional quantum systems where the density matrix is represented by a (1+1)-dimensional tensor network, which is well-suited for TRG computations.
We have tested our method on the isotropic classical Ising model, though the approach can be directly applied to more general systems with quantum Hamiltonians.
The entanglement entropy at the critical point shows logarithmic scaling with the proper central charge as expected.

The additional computational cost for evaluating the entanglement entropy is $O(D^{\max (2h+2, 3h)})$ in two-dimensional tensor networks, where $D$ is the bond dimension of the tensor network and $h$ is the Hamming weight of the subsystem size $\ell$ expressed in binary form.
To keep the computational cost low, we took a single interval of length $2^m$ or $2^m + 2^q$ with $h=1$ or 2 as a subsystem.
Our method can be straightforwardly generalized to higher-dimensional systems as long as we take a hyperrectangle as a subsystem.
In this case, Fig.~\ref{fig.rho_example} and Fig.~\ref{fig.general_rho} has $(d-1)$-dimensional isometries.
The computational cost is reasonable if the Hamming weights of edges of the hyperrectangle are sufficiently small.

The entanglement entropy would be an interesting quantity to study field theory from various aspects such as phase structure, holography and quantum information.
Our method will serve as a useful tool for this purpose.

\section*{Acknowledgement}
We would like to thank Harunobu Fujimura, Tomoya Hayata, and Donghoon Kim for the useful comments and discussions.
This work was partially supported by JSPS KAKENHI Grant Number
21K03531,	
21K03537,	
22H01222,	
22H05251,	
23H00112,	
23K22493,	
and 25K07280.	
This work was supported by JST SPRING, Grant Number JPMJSP2135.	

\appendix
\section{Notations}
\label{sec:notation}
\begin{figure}[H]
	\centering
	\begin{minipage}[b]{0.48\linewidth}
		\centering
		\includegraphics[keepaspectratio, scale=0.7]{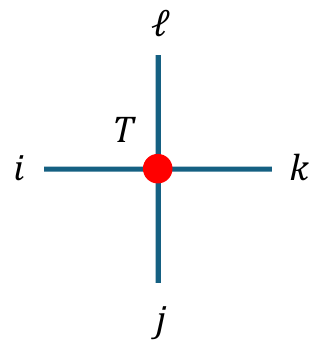}
		\caption{Tensor $T_{ijk\ell}$.}
		\label{fig.Tensor}
	\end{minipage}
	\centering
	\begin{minipage}[b]{0.48\linewidth}
		\centering
		\includegraphics[keepaspectratio, scale=0.7]{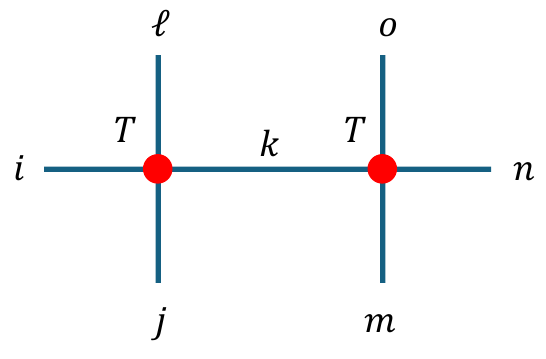}
		\caption{Tensor $(TT)_{ijmno\ell} = \sum_k T_{ijkl} T_{kmno}$.}
		\label{fig.Tensor_cont}
	\end{minipage}
\end{figure}

A rank $n$ tensor $T$ is denoted as $T_{i_1 i_2 \cdots i_n}$, where $i_k$ $(k=1,2,\cdots,n)$ runs from $1$ to $D$.
Throughout this paper, tensor diagram is used to represent a product of tensors graphically.
In the tensor diagram, $T_{i_1 i_2 \cdots i_n}$ is expressed as a symbol with $n$ lines attached to it, where each line corresponds to an index of the tensor.
Internal lines between two tensors correspond to the contraction of indices.
For example, a tensor $T_{ijk\ell}$ is illustrated as shown in Fig.~\ref{fig.Tensor}.
The contraction of two tensors $T_{ijkl}$ and $T_{kmno}$ is illustrated as shown in Fig.~\ref{fig.Tensor_cont}.

\bibliographystyle{JHEP}
\bibliography{hotrg_ee}

\end{document}